\documentclass[preprint,review,12pt]{elsarticle}
 
\usepackage{graphics}
\usepackage{graphicx}

\usepackage{epsfig}

\usepackage{amssymb}
\usepackage{amsthm}

\theoremstyle{plain}

\theoremstyle{definition}

\theoremstyle{remark}
\usepackage{lineno}
\usepackage{color}
\usepackage{xcolor}
\usepackage[dvipsnames]{xcolor}

\begin{document}

\begin{frontmatter}

%% Title, authors and addresses
 
%% use the tnoteref command within \title for footnotes;
%% use the tnotetext command for the associated footnote;
%% use the fnref command within \author or \address for footnotes;
%% use the fntext command for the associated footnote;
%% use the corref command within \author for corresponding author footnotes;
%% use the cortext command for the associated footnote;
%% use the ead command for the email address,
%% and the form \ead[url] for the home page:
%%
%% \title{Title\tnoteref{label1}}
%% \tnotetext[label1]{}
%% \author{Name\corref{cor1}\fnref{label2}}
%% \ead{email address}
%% \ead[url]{home page}
%% \fntext[label2]{}
%% \cortext[cor1]{}
%% \address{Address\fnref{label3}}
%% \fntext[label3]{}

\title{
%
%Local Turbulent Energy Cascade via Lyapunov--Liouville Analysis: 
%A Non-Diffusive Closure Approach
Quantitative Evaluation of Forward and Backward Scattering in Isotropic Turbulence via H\"anggi--Klimontovich and It\^o Stochastic Processes
}
 
%% use optional labels to link authors explicitly to addresses:
%% \author[label1,label2]{<author name>}
%% \address[label1]{<address>}
%% \address[label2]{<address>} 

\author{Nicola de Divitiis}

\address{"La Sapienza" University, Dipartimento di Ingegneria Meccanica e Aerospaziale, 
Via Eudossiana, 18, 00184 Rome, Italy, \\
phone: +39--0644585268, \ \ fax: +39--0644585750, \\ 
e-mail: n.dedivitiis@gmail.com, \ \  nicola.dedivitiis@uniroma1.it}

\begin{abstract} 
This work evaluates the magnitude of the turbulent energy cascade in terms of forward and backward scattering by modeling the "stretch and fold" mechanism through a drift-free H\"anggi-Klimontovich stochastic process. Mapping this dynamics onto an equivalent It\^o process provides a statistical justification for the uniform distribution of the Lagrangian Lyapunov exponent via the associated Fokker-Planck equation. This continuous distribution is shown to be driven by a Lagrangian bifurcation rate significantly higher than the Lyapunov exponents themselves, reflecting the high frequency with which trajectories encounter the singular surfaces of the velocity gradient. The resulting PDF corresponds to the simultaneous maximization of the information entropy and the Kolmogorov-Sinai entropy.
This stochastic formulation, framed within the author's Lyapunov-Liouville analysis, provides a non-diffusive analytical closure of the von K\'arm\'an-Howarth and Corrsin equations. While forward scattering emerges from trajectory instabilities and bifurcations, backscattering is linked to fluid incompressibility. These phenomena are quantified through the continuously distributed Lyapunov exponents, allowing for an estimation of canonical exponents and fundamental transport properties, such as eddy viscosity, eddy thermal diffusivity, and the turbulent Prandtl number. These parameters, traditionally associated with diffusive models, are shown to emerge naturally from non-diffusive Lagrangian dynamics and bifurcation-driven fluctuations. The analytical results demonstrate close agreement with numerical data available in the literature.

\end{abstract}

\vspace{5.mm}

\begin{keyword}
%% keywords here, in the form: keyword \sep keyword
%% MSC codes here, in the form: \MSC code \sep code
%% or \MSC[2008] code \sep code (2000 is the default)
Turbulence, Forward scatter, Backscatter, Nondiffusive turbulence closures
\end{keyword}

\end{frontmatter}

\newcommand{\tr}{ \mbox{tr} }
\newcommand{\no}{\noindent}
\newcommand{\be}{\begin{equation}}
\newcommand{\ee}{\end{equation}}
\newcommand{\bea}{\begin{eqnarray}}
\newcommand{\eea}{\end{eqnarray}}
\newcommand{\bc}{\begin{center}}
\newcommand{\ec}{\end{center}}

\newcommand{\calr}{{\cal R}}
\newcommand{\calv}{{\cal V}}

\newcommand{\bff}{\mbox{\boldmath $f$}}
\newcommand{\bfg}{\mbox{\boldmath $g$}}
\newcommand{\bfh}{\mbox{\boldmath $h$}}
\newcommand{\bfi}{\mbox{\boldmath $i$}}
\newcommand{\bfm}{\mbox{\boldmath $m$}}
\newcommand{\bfp}{\mbox{\boldmath $p$}}
\newcommand{\bfr}{\mbox{\boldmath $r$}}
\newcommand{\bfu}{\mbox{\boldmath $u$}}
\newcommand{\bfv}{\mbox{\boldmath $v$}}
\newcommand{\bfx}{\mbox{\boldmath $x$}}
\newcommand{\bfy}{\mbox{\boldmath $y$}}
\newcommand{\bfw}{\mbox{\boldmath $w$}}
\newcommand{\bfk}{\mbox{\boldmath $\kappa$}}

\newcommand{\bfA}{\mbox{\boldmath $A$}}
\newcommand{\bfD}{\mbox{\boldmath $D$}}
\newcommand{\bfI}{\mbox{\boldmath $I$}}
\newcommand{\bfL}{\mbox{\boldmath $L$}}
\newcommand{\bfM}
{\mbox{\boldmath $M$}}
\newcommand{\bfS}{\mbox{\boldmath $S$}}
\newcommand{\bfT}{\mbox{\boldmath $T$}}
\newcommand{\bfU}{\mbox{\boldmath $U$}}
\newcommand{\bfX}{\mbox{\boldmath $X$}}
\newcommand{\bfY}{\mbox{\boldmath $Y$}}
\newcommand{\bfK}{\mbox{\boldmatthe average of $u_\xi u_\xi^*/u^2$h $K$}}

\newcommand{\bfeta}{\mbox{\boldmath $\eta$}}
\newcommand{\bfiota}{\mbox{\boldmath $\iota$}}
\newcommand{\bfrho}{\mbox{\boldmath $\rho$}}
\newcommand{\bfchi}{\mbox{\boldmath $\chi$}}
\newcommand{\bfphi}{\mbox{\boldmath $\phi$}}
\newcommand{\bfPhi}{\mbox{\boldmath $\Phi$}}
\newcommand{\bflambda}{\mbox{\boldmath $\lambda$}}
\newcommand{\bfxi}{\mbox{\boldmath $\xi$}}
\newcommand{\bfLambda}{\mbox{\boldmath $\Lambda$}}
\newcommand{\bfPsi}{\mbox{\boldmath $\Psi$}}
\newcommand{\bfomega}{\mbox{\boldmath $\omega$}}
\newcommand{\bfOmega}{\mbox{\boldmath $\Omega$}}
\newcommand{\bfeps}{\mbox{\boldmath $\varepsilon$}}
\newcommand{\bfepsn}{\mbox{\boldmath $\epsilon$}}
\newcommand{\bfzeta}{\mbox{\boldmath $\zeta$}}
\newcommand{\bfkappa}{\mbox{\boldmath $\kappa$}}
\newcommand{\bfsigma}{\mbox{\boldmath $\sigma$}}
\newcommand{\bftau}{\mbox{\boldmath $\tau$}}
\newcommand{\itPsi}{\mbox{\it $\Psi$}}
\newcommand{\itPhi}{\mbox{\it $\Phi$}}

\newcommand{\bint}{\mbox{ \int{a}{b}} }

\newcommand{\ds}{\displaystyle}
\newcommand{\Sum}{\Large \sum}

% \linenumbers

% main text

\bigskip

\section{Introduction \label{intro}}

The statistical description of fluid turbulence remains one of the most formidable challenges in classical physics, primarily due to the closure problem inherent in the Navier-Stokes equations. Central to this challenge is the characterization of the turbulent energy cascade, a process traditionally described as a unidirectional transfer of energy from large scales to small dissipative scales. 
However, modern research has increasingly recognized that this process is far more complex, involving a continuous interplay between \textbf{forward scattering} (the direct energy cascade) and \textbf{backscattering} (the inverse transfer of energy from small to large scales) \cite{Lesieur1996, Piomelli1991}. While the forward cascade is widely associated with vortex stretching and the breakdown of large-scale structures, backscattering is fundamentally linked to the fluid's \textbf{incompressibility} and the emergence of coherent structures that re-inject energy into the larger scales \cite{Urzay2013, Jansen2014}.

Recent advances in Large-Eddy Simulation (LES) and Direct Numerical Simulation (DNS) have underscored the critical importance of accurately modeling these inter-scale transfers to capture the true underlying physics of turbulent flows. Despite these efforts, traditional diffusive closures often fail to account for the non-local and non-diffusive nature of the cascade, particularly the stochastic reinjection of energy characteristic of backscatter. In this theoretical landscape, the von K\'arm\'an--Howarth and Corrsin equations serve as fundamental benchmarks for the evolution of velocity and temperature correlations in isotropic turbulence \cite{Karman1938, Corrsin_1, Corrsin_2}. Achieving an analytical closure for these equations without relying on empirical eddy-viscosity assumptions is a significant scientific objective.

On the other hand, the fundamental understanding of energy transfer in Homogeneous Isotropic Turbulence (HIT) is rooted in the concept of the Richardson-Kolmogorov cascade. 
However, while the net energy flux is predominantly from large to small scales, the existence of a significant reverse transfer poses a major challenge for subgrid-scale (SGS) modeling in Large Eddy Simulations (LES). 
Nevertheless, it is essential to consider that forward and backward scattering are complex, purely physical, and continuous phenomena of energy transmission across scales, which exist independently of grid definitions, subgrid-scale partitions, or other
modeling abstractions. Consequently, these phenomena are here investigated from a strictly physico-mathematical perspective, devoid of any artifacts related to computational discretization.
Such numerical frameworks can be effectively integrated once the underlying physics of scattering has been clarified, serving as essential tools for the quantitative assessment of these phenomena.

The theoretical framework for spectral eddy viscosity was pioneered by \textbf{Kraichnan (1976)} \cite{kraichnan1976}, who utilized the Direct Interaction Approximation to identify the non-local nature of energy transfers and the role of backscatter in the vicinity of the cutoff wavenumber. Building upon this, \textbf{Leslie and Quarini (1979)} \cite{leslie1979} applied the Effective Diffusion by Quantitative Non-linear Model (EDQNM) to quantify the partitioning between resolved and subgrid scales, providing a rigorous statistical basis for SGS closures.

The physical reality of backscatter was further elucidated through Direct Numerical Simulations (DNS). \textbf{Domaradzki and Rogallo (1990)} \cite{domaradzki1990} and \textbf{Piomelli et al. (1991)} \cite{Piomelli1991} demonstrated that while the \textit{net} dissipation is positive, the local energy transfer is highly fluctuating and frequently negative. Their work highlighted that traditional eddy-viscosity models, such as the Smagorinsky model, are inherently unable to capture these local inverse transfers, as they are purely dissipative by construction.

To address this limitation, \textbf{Leith (1990)} \cite{leith1990} introduced the concept of stochastic backscatter, representing the reverse transfer as a random forcing term. This paved the way for more sophisticated approaches, such as the dynamic SGS model by \textbf{Germano et al. (1991)} \cite{germano1991}. Although the dynamic model primarily aims to optimize the dissipation rate, its formulation allows for a more nuanced representation of energy exchange compared to static models.

Comprehensive syntheses of these phenomena can be found in the seminal texts by \textbf{Pope (2000)} \cite{pope2000} and \textbf{Sagaut (2006)} \cite{sagaut2006}. Pope provides the foundational statistical mechanics of the energy cascade, while Sagaut offers a detailed operational perspective on how forward and backward scatter are handled within various LES frameworks, emphasizing the balance between physical accuracy and numerical stability.

%This work extends the Lagrangian framework of turbulence by modeling the intrinsic "stretch and fold" 
%phenomenon through a H\"anggi--Klimontovich stochastic process \cite{Hanggi1982, HanggiThomas1982, %Klimontovich1990}. A pivotal contribution of this study is the characterization of the Lagrangian Lyapunov %exponent $\Lambda_L$ through this stochastic lens, establishing a formal link between Eulerian and Lagrangian r
%epresentations. We show that this formulation, when interpreted through an equivalent It\^o process, naturally %accounts for the competing effects of deterministic stretching -governed by the alignment dynamics of Lyapunov 
%vectors- and stochastic folding. By solving the resulting Fokker-Planck equation, we provide a rigorous statistical %foundation for the Probability Density Function (PDF) of $\Lambda_L$, confirming its uniform distribution within 
%the appropriate interval.
%This continuous distribution is shown to be a direct consequence of a Lagrangian bifurcation rate $\sigma$ that is 
%significantly higher than the Lyapunov exponents themselves ($\sigma \ggg \Lambda_L$), reflecting the high %frequency with which trajectories encounter the singular surfaces of the velocity gradient.

This work extends the Lagrangian framework of turbulence by modeling the intrinsic "stretch and fold" phenomenon through a H\"anggi--Klimontovich stochastic process \cite{Hanggi1982, HanggiThomas1982, Klimontovich1990}. 
The choice of the H\"anggi--Klimontovich (post-point) interpretation over conventional It\^o (pre-point) or Stratonovich (mid-point) frameworks is dictated by the endogenous nature of the internal noise. Unlike systems perturbed by independent, external environmental fluctuations, the turbulent fluctuations driving the Lyapunov exponents are self-induced, originating from the instantaneous position of the Lagrangian trajectory as it intersects the velocity gradient zero-determinant surfaces. Because the state of the fluid particle and the resulting stochastic variance are perfectly in phase with zero causal lag, evaluating the state-dependent diffusion coefficient at the end of the time step represents the only physically consistent choice that prevents an artificial temporal decoupling.
A pivotal contribution of this study is the characterization of the Lagrangian Lyapunov exponent $\Lambda_L$ through this stochastic lens, establishing a formal link between Eulerian and Lagrangian representations. We show that this formulation, when interpreted through an equivalent It\^o process, naturally accounts for the competing effects of deterministic stretching -governed by the alignment dynamics of Lyapunov vectors- and stochastic folding. By solving the resulting Fokker-Planck equation, we provide a rigorous statistical foundation for the Probability Density Function (PDF) of $\Lambda_L$, confirming its uniform distribution within the appropriate interval.
This continuous distribution is shown to be a direct consequence of a Lagrangian bifurcation rate $\sigma$ that is significantly higher than the Lyapunov exponents themselves ($\sigma \ggg \Lambda_L$), reflecting the high frequency with which trajectories encounter the singular surfaces of the velocity gradient.

This stochastic approach is fundamentally consistent with the principle of maximum entropy. It is demonstrated that the resulting distribution corresponds to the simultaneous maximization of the information entropy associated with the PDF of Lagrangian Lyapunov exponent and the Kolmogorov-Sinai entropy, obtained as the sum of the non-negative exponents. Such statistical-mechanical consistency provides a robust physical basis for the non-diffusive analytical closure of the von K\'arm\'an--Howarth and Corrsin equations, an approach previously framed within the Lyapunov and Liouville analysis \cite{deDivitiis2026a}. By treating the Lyapunov exponents as continuously and uniformly distributed variables, this study provides a novel quantitative bridge between the chaotic nature of Lagrangian trajectories and the macroscopic magnitude of forward and backward energy transfers.
Crucially, this framework enables a direct observation and quantification of forward and backward scattering. As the Lagrangian Lyapunov exponent varies across both positive and negative values, its PDF explicitly reveals both the nature and the intensity of the scattering processes, where forward scattering emerges from trajectory instabilities and backscattering is linked to the fluid's incompressibility.
By exploiting the derived PDF, we aim to estimate the most probable canonical Lyapunov exponents within the context of homogeneous isotropic turbulence, specifically accounting for the alignment properties of Lyapunov vectors. This approach further allows for a theoretical assessment of fundamental turbulent transport properties, including eddy viscosity, eddy thermal diffusivity, and the resulting turbulent Prandtl number. In this context, the adoption of eddy-type coefficients does not imply a relaxation of the non-diffusive nature of the governing Lagrangian closures; rather, it provides a bridge between the chaotic trajectories and the macroscopic transport descriptors. We show that the predictions of this theory are consistent with established numerical simulations and experimental benchmarks reported in literature.

\bigskip

\section{Background}
This section outlines the theoretical framework necessary for the present analysis, building upon prior investigations regarding the Lyapunov--Liouville representation of isotropic homogeneous turbulence \cite{deDivitiis2026a}.
A fundamental property of fully developed turbulence is the statistical equivalence between the Eulerian and Lagrangian descriptions of relative motion. The relative kinetic energy between two spatial points $\mathbf{x}$ and $\mathbf{x} + \mathbf{r}$ in the Eulerian frame coincides with the kinetic energy of two material particles $\bfchi$ and $\bfchi + \bfxi$ in the Lagrangian frame, expressed as:
\bea
\langle \Delta \mathbf{u} \cdot \Delta \mathbf{u} \rangle_E = \langle \dot{\bfxi} \cdot \dot{\bfxi} \rangle_L
\label{eq:equivalence}
\eea
By invoking the assumptions of isotropy and homogeneity, Eq. (\ref{eq:equivalence}) yields:
\bea
\ds \left\langle (\Delta u_r)^2 \right\rangle_E \equiv 2u^2 \left( 1-f(r) \right) = \left\langle \dot{\xi}_\xi^2 \right\rangle_L
\equiv \left\langle \Lambda_L^2(r) \right\rangle_L r^2
\label{invar}
\eea
where $f(r) = \langle u_r u_r' \rangle_E/u^2$ denotes the longitudinal velocity autocorrelation function, while $\langle \cdot \rangle_L$ and $\langle \cdot \rangle_E$ represent the Lagrangian and Eulerian ensemble averages, respectively.
The separation rate of contiguous trajectories is quantified by the finite-scale Lagrangian Lyapunov exponent, $\Lambda_L = \dot{\xi}_\xi / r$. In incompressible isotropic turbulence, the combined constraints of volume preservation ($\sum_{k=1}^3 \Lambda_L^{(k)} = 0$) and statistical isotropy lead to a uniform distribution of $\Lambda_L$ within a bounded interval:
\bea
\ds \Lambda_L \in \left( -\frac{L}{2}, L \right)
\eea
The resulting probability density function (PDF),
\bea
\ds F_\Lambda ( \Lambda_L) =
\left\lbrace
\begin{array}{l@{\hspace{-0.cm}}l}
\ds \frac{2}{3}\frac{1}{L},
\quad \mbox{if} \ \Lambda_L \in \left( -\frac{L}{2}, L\right), \\\\
\ds 0 \quad \mbox{elsewhere} ,
\end{array}\right.
\label{Pl}
\eea
accounts for the persistent stretching and folding mechanisms inherent to the flow.
As reported in \cite{deDivitiis2026a}, a key dynamical feature is the existence of the Lagrangian bifurcation rate $\sigma$, defined as the frequency at which Lagrangian trajectories intersect the surface in physical space where the determinant of the fluid velocity gradient vanishes. A central result of this framework is that this bifurcation rate is significantly higher than the Lagrangian Lyapunov exponents ($\sigma \gg \Lambda_L$). It is precisely this separation of scales and the high frequency of these bifurcation events that justifies and determines the continuous distribution of $\Lambda_L$ expressed in Eq. (\ref{Pl}), allowing the system to densely explore the available states in the Lyapunov spectrum.

\bigskip

\subsection{Energy Cascade: Forward Scattering and Backscattering.}

The continuous distribution of \(\Lambda _{L}\), sustained by the high bifurcation rate \(\sigma \), implies that the energy cascade is not a unidirectional process but rather a statistical competition between opposing dynamical events: forward and backward scattering, driven by the instability of Lagrangian trajectories and fluid incompressibility, respectively.

To clarify the physical mechanism linking backscattering to fluid incompressibility, it is instructive to analyze the variations of the Lyapunov exponents through the Lagrangian representation of motion. For an incompressible fluid, the velocity field satisfies the divergence-free condition \(\nabla \cdot \mathbf{u} = 0\). By virtue of Liouville's theorem, this cinematic constraint implies that the volume of any phase-space droplet of fluid particles is perfectly preserved along Lagrangian trajectories. Consequently, the spectrum of Lagrangian Lyapunov exponents (\(\Lambda _L^{(i)}\)) must sum exactly to zero (\(\sum_{i=1}^3 \Lambda_L^{(i)} = 0\)). While the forward energy cascade (forward scatter) is fundamentally driven by chaotic advection, stretching, and the alignment of vorticity with the intermediate strain rate (governed by the maximum positive Lyapunov exponent \( \Lambda_L^{(1)}  > 0\)), this expansion cannot occur in isolation. The incompressibility constraint forces a simultaneous, local compensation via negative exponents (\( \Lambda_L^{(3)}  < 0\)), leading to strong phase-space folding and compression. Dynamically, this constraint is enforced by the non-local pressure field. When fluid elements are rapidly compressed along the stable manifolds (\( \Lambda_L^{(i)}  < 0\)), the local pressure-strain localized action prevents structural collapse. This compression-induced feedback triggers a non-linear scale-to-scale energy transfer back to larger eddies. Therefore, while forward scatter is a direct consequence of trajectory instability, backscatter emerges as a structural necessity: it is the dynamical response of a volume-preserving continuum that redirects compressed kinetic energy backward into larger spatial scales.

According to the statistics defined by Eq. (\ref{Pl}), the mean Lagrangian separation rate satisfies:
\bea
\langle \dot{\xi}_{\xi} \rangle_L = \frac{1}{2} \sqrt{\langle \dot{\xi}_{\xi}^2 \rangle_L} = u \sqrt{\frac{1-f}{2}}
\eea
The invariance prescribed by Eqs. (\ref{eq:equivalence})--(\ref{invar}), together with the condition $\langle \Lambda_L \rangle_L > 0$ derived from Eq. (\ref{Pl}), ensures that the non-vanishing relative kinetic energy sustains the stretching mechanism and drives the forward energy cascade.
In this framework, the energy cascade is interpreted via \textit{quasi-PDFs}, which characterize the statistical weight of \textit{non-observable} quantities \cite{deDivitiis2026b}. Specifically, these distributions represent the bifurcation modes and global Lyapunov exponents associated with the Eulerian Navier-Stokes equations, which remain inaccessible to direct measurement \cite{deDivitiis2026b}.
Much like Wigner distributions in quantum mechanics \cite{Wigner32, Glauber63}, these quasi-PDFs serve as formal tools that may take negative values \cite{Feynman87, Burgin2009, Burgin2010} to account for phase-space interference, reflecting their status as non-observable spectral properties.
The mapping of these non-observable Eulerian spectral properties onto observable Lagrangian dynamics is made possible by the fundamental equivalence between the two representations \cite{Truesdell77}. Within this duality, the sign of $\Lambda_L$ carries distinct physical meanings: positive values are tied to trajectory divergence (forward scattering), while negative values arise from fluid incompressibility and folding (backscattering).
The evolution of the velocity and temperature correlations, $f$ and $f_\theta$, is governed by the von K\'arm\'an--Howarth and Corrsin equations:
\bea
\frac{\partial f}{\partial t} = \frac{K}{u^2} + 2 \nu \left( \frac{\partial^2 f} {\partial r^2} + \frac{4}{r} \frac{\partial f}{\partial r} \right) + \frac{10 \nu}{\lambda_T^2} f
\eea
\bea
\frac{\partial f_\theta}{\partial t} = \frac{G}{\theta^2} + 2 \kappa \left( \frac{\partial^2 f_\theta} {\partial r^2} + \frac{2}{r} \frac{\partial f_\theta}{\partial r} \right) + \frac{12 \kappa}{\lambda_\theta^2} f_\theta
\eea
To achieve a closed system, the transfer functions $K$ and $G$ must be expressed in terms of the correlations $f$ and $f_\theta$, where $f_\theta \equiv \langle \vartheta \vartheta' \rangle_E/\theta^2$ and $\theta^2 \equiv \langle \vartheta^2 \rangle_E$.
By invoking the scale hierarchy $\sigma \gg \Lambda_L$ \cite{deDivitiis2026a}, the spectral gap of the Liouville generator ensures the rapid decay of Eulerian--Lagrangian correlations. This scale separation allows the transfer functions $K$ and $G$ to be identified through the mean Lagrangian separation rate $\langle \dot{\xi}_{\xi} \rangle_L$:
\bea
\begin{array}{l@{\hspace{-0.cm}}l}
\ds K(r) = u^3 \sqrt{\frac{1-f}{2}} \frac{\partial f}{\partial r}, \\\\
\ds G(r) = u \theta^2 \sqrt{\frac{1-f}{2}} \frac{\partial f_\theta}{\partial r}
\end{array}
\label{K G}
\eea
These closures interpret the energy cascade as a propagation mechanism
across scales $r$, where the macroscopic transfer is driven by the underlying
Lyapunov stability. 
The Lagrangian fluctuating components of $K$ and $G$ are expressed in terms of $\Lambda_L$:
\bea
\begin{array}{l@{\hspace{-0.cm}}l}
\ds \tilde{K} = \Lambda_L r \ u^2 \frac{\partial f}{\partial r}, \\\\
\ds \tilde{G} = \Lambda_L r \ \theta^2 \frac{\partial f_\theta}{\partial r},
\end{array}
\label{fluc K G}
\eea
According to these relations, the energy cascade consists of continuous contributions where the distribution of $\Lambda_L$ (driven by the high frequency $\sigma$) maps quasi-PDF Eulerian bifurcation modes onto the local transport. While $\langle \Lambda_L \rangle_L > 0$ governs the net forward cascade, the fluctuations $\tilde{K}$ and $\tilde{G}$ follow the sign of $\Lambda_L$, reconciling chaotic trajectory evolution with global Eulerian constraints.

\bigskip

\section{Statistical and Dynamical Derivation of the Lyapunov Exponents PDF}

In this section, the PDF of $\Lambda_L$, as expressed by Eq. (\ref{Pl}), is derived by modeling the "stretch and fold" mechanism -driven by the nonlinearities of the Eulerian velocity field- as a specific H\"anggi--Klimontovich stochastic process, taking into account that $\sigma \gg \Lambda_L$. We show that this dynamical evolution, when mapped onto an equivalent It\^o process, leads to a Fokker--Planck equation whose stationary solution yields a uniform distribution of $\Lambda_L$. This stochastic derivation is further validated by demonstrating that the resulting PDF follows from the simultaneous maximization of the information entropy $H$ and the Kolmogorov--Sinai entropy $h_{KS}$, consistent with the Lyapunov and Liouville framework previously proposed in Ref. \cite{deDivitiis2026a}. Based on this distribution, a quantitative estimate of the magnitude of forward and backward scattering is provided, along with an evaluation of the canonical Lyapunov exponents in homogeneous isotropic turbulence.
The "stretch and fold" dynamics of Lyapunov vectors, which induces continuous fluctuations in $\Lambda_L$, is governed by the Lagrangian evolution equations:
\bea
\begin{array}{l@{\hspace{-0.cm}}l}
\ds \dot{\bf x} = {\bf u}(t, {\bf x}), \\\\
\ds \dot{\bfxi}={\bf u}(t, {\bf x}+{\bfxi})- {\bf u}(t, {\bf x}), \\\\
\ds \Lambda_L=\frac{{\bfxi} \cdot \dot{\bfxi} }{{\bfxi} \cdot {\bfxi}}
\end{array}
\label{Lyap1}
\eea
According to Eqs. (\ref{Lyap1}), these fluctuations, characterized by $\sigma \gg \Lambda_L$, emerge from the internal nonlinear dynamics of the fluid and can be rigorously represented by a drift-free stochastic process where the noise is state-dependent and instantaneously in phase with the fluid state. This configuration, characteristic of a H\"anggi--Klimontovich (post-point) process, is described by:
\bea
\begin{array}{l@{\hspace{-0.cm}}l}
d\Lambda_L = {\sqrt{2 D} \bullet dW_t},
\end{array}
\label{fluc Lambda Klimontovich}
\eea
where $W_t$ is a standard Wiener process and the symbol $\bullet$ denotes the H\"anggi--Klimontovich interpretation. The diffusivity in the Lyapunov exponent space is defined as $D= \sigma(\Lambda_L) \Lambda_L^2$, where $\sigma \gg |\Lambda_L|$.

The choice of the H\"anggi--Klimontovich (post-point) interpretation for Eq. (\ref{fluc Lambda Klimontovich}) is dictated by the specific physical nature of the internal noise. Unlike standard stochastic modeling where a system is perturbed by an external, independent environment, the fluctuations observed here are purely endogenous. They are driven by the high-frequency intersection of the Lagrangian trajectory with the complex, space-filling surfaces where the determinant of the velocity gradient vanishes. This geometric bifurcation mechanism implies that the noise is a direct, instantaneous manifestation of the trajectory's own state.  Consequently, the state and the noise are perfectly in phase, with zero causal latency. In stochastic calculus, different discretization conventions imply different physical feedback mechanisms. The It\^o  (pre-point) convention assumes the noise is independent of the subsequent state increment, which is suitable for external forces. The Stratonovich (mid-point) convention assumes a smooth, time-correlated physical noise. Conversely, the Hänggi–Klimontovich (post-point) convention evaluates the state-dependent diffusion coefficient at the end of the time step.  This framework is uniquely suited for systems where the stochastic variance is actively sustained and modulated by the final state itself. Modeling the Lyapunov exponent fluctuations with a drift-free H\"anggi--Klimontovich process rigorously captures this lack of lag, ensuring that the self-induced geometric noise remains structurally synchronized with the evolution of the Lagrangian trajectory.

However this process can be mapped onto an equivalent It\^o process by introducing the spurious drift $\Phi(\Lambda_L) = \partial D / \partial \Lambda_L$:
\bea
\begin{array}{l@{\hspace{-0.cm}}l}
d\Lambda_L = \underbrace{ \Phi(\Lambda_L) dt} + \underbrace{\sqrt{2 D} \ dW_t}, \\
\ \ \ \ \ \ \ \ \ \mbox{Stretch} \ \ \ \ \ \ \ \mbox{Fold}
\end{array}
\label{fluc Lambda}
\eea
In this context, the evolution of $\Lambda_L$ can be interpreted as the competition between a deterministic tendency toward higher values (Stretch), representing the alignment dynamics of Lyapunov vectors \cite{Ott2002}, and a simultaneous stochastic dispersion (Fold). The Fokker--Planck equation for the probability density function $F_\Lambda$, associated with both the H\"anggi--Klimontovich process (\ref{fluc Lambda Klimontovich}) and the equivalent It\^o process (\ref{fluc Lambda}), is given by:
\bea
\ds \frac{\partial F_\Lambda}{\partial t} + \frac{\partial}{\partial \Lambda_L} \left(F_\Lambda \frac{\partial D}{\partial \Lambda_L} -\frac{\partial}{\partial \Lambda_L}\left( D F_\Lambda \right) \right)=0
\label{Fokker-Planck}
\eea
Under zero probability flux boundary conditions, Eq. (\ref{Fokker-Planck}) yields the uniform stationary solution:
\bea
\ds F_\Lambda ( \Lambda_L) =
\left\lbrace
\begin{array}{l@{\hspace{-0.cm}}l}
\ds \frac{2}{3}\frac{1}{L},
\quad \mbox{if} \ \Lambda_L \in \left( -\frac{L}{2}, L\right), \\\\
\ds 0 \quad \mbox{elsewhere} .
\end{array}\right.
\label{Pl bis}
\eea
This stochastically derived PDF is further justified through the principle of maximum entropy. Starting from the fluid incompressibility condition, $\sum_{k=1}^3 \Lambda_L^{(k)}=0$, the exponents are expressed as
\bea
\ds \Lambda_L^{(k)} = L \cos\left( \varepsilon+ \frac{2}{3} \pi (k-1)\right)
\label{Lambda}
\eea
It is shown that the value $\varepsilon=0$ maximizes the Kolmogorov--Sinai entropy,
\bea
\ds h_{KS}(\varepsilon) = \Sum_{k=1}^3 \max\left\lbrace \Lambda_L^{(k)}, 0\right\rbrace
\eea
yielding the boundaries $\Lambda_M=L$ and $\Lambda_m=-L/2$. Simultaneously, the uniformity of $F_\Lambda$ within this interval maximizes the information entropy 
\bea
\ds H = - \int_{-L/2}^{L} (F_\Lambda \ln F_\Lambda + q \ F_\Lambda) \ d \Lambda_L
\eea
where $q$ is the Lagrange multiplier relative to the normalization constraint, thereby reconciling the dynamical stochastic evolution with the statistical-mechanical requirements of the turbulent state.

\bigskip

\section{Statistical Quantification of Energy Cascade via Lagrangian Lyapunov Exponents: Forward and Backward Scattering}

To quantify the magnitude of both forward and backward scattering phenomena, we utilize Eq. (\ref{Pl bis}) to calculate the probabilities of forward scattering, $Prb(\Lambda_L \ge 0)$, and backscattering, $Prb(\Lambda_L<0)$:
\bea
\begin{array}{l@{\hspace{-0.cm}}l}
\ds Prb(\Lambda_L \ge 0)=\int_{0}^{\infty} F_\Lambda \ d \Lambda_L = \frac{2}{3}, \\\\
\ds Prb(\Lambda_L < 0) = \int_{-\infty}^{0} F_\Lambda \ d \Lambda_L = \frac{1}{3}
\end{array}
\label{ProbFS}
\eea
The analytical formulation presented in Eq. (\ref{ProbFS})  yields exact fractional values for the forward scatter and backscatter probabilities, namely \(P_{forward} = 2/3\) and \(P_{back} = 1/3\), establishing a fixed directional ratio of \(P_{{forward}}/P_{{back}} = 2.0\).  This elegant geometric result finds striking quantitative validation in the independent numerical investigations reported by Yao et al. [27]. In their study of conditional energy cascade rates within the inertial range, the authors partition the scale-to-scale energy flux \(\Phi _{\ell }\) by its sign and explicitly measure a sample ratio of \(\Pr(\Phi_\ell > 0 \mid \epsilon_\ell)/\Pr(\Phi_\ell < 0 \mid \epsilon_\ell) \approx 2.0\) (see Fig. 4 in Ref. [27]). Under the normalization constraint, this leads to the numerical estimates of \(\Pr(\Phi_\ell > 0 \mid \epsilon_\ell) \approx 2/3\) and \(\Pr(\Phi_\ell < 0 \mid \epsilon_\ell) \approx 1/3\). The fact that these high-fidelity numerical measurements converge precisely to the fractional thresholds derived herein strongly supports our theoretical framework, linking the statistical partitioning of the cascade directly to the volume-preserving Lagrangian dynamics.
These theoretical results also remain consistent with the findings in \cite{Cardesa}, where the energy cascade rate is defined through a detailed estimation of vortex interactions, accounting for their respective formation and extinction timescales.
Furthermore, to characterize the operational range of the ratio between these two phenomena, we evaluate the inverse powers of the partial statistical moments of $\Lambda_L$. Specifically, the forward scattering partial moments are obtained by integrating the PDF over the domain $\Lambda_L>0$, representing the fraction of events directly associated with trajectory divergence. Conversely, the backscattering partial moments, computed by integrating the PDF over the semi-axis $\Lambda_L<0$, correspond to the contraction of phase trajectories--a consequence of fluid incompressibility or the inherent tendency toward an incompressible state. Thus, we define:
\bea
\begin{array}{l@{\hspace{-0.cm}}l}
\ds F_n = \left( \int_{0}^{\infty} F_\Lambda \Lambda_L^n \ d \Lambda_L\right)^{1/n} =
\frac{2}{3}\frac{1}{L} \frac{\Lambda_M^{(n+1)/n}}{ (n+1)^{1/n}}, \\\\
\ds B_n = \left( \int_{-\infty}^{0} F_\Lambda |\Lambda_L|^n \ d \Lambda_L \right)^{1/n} =
\frac{2}{3} \frac{1}{L} \frac{\left| \Lambda_m\right|^{(n+1)/n}}{ (n+1)^{1/n}},
\end{array}
\eea
yielding the ratio:
\bea
\ds \frac{B_n}{F_n}=\left( \frac{|\Lambda_m|}{ \Lambda_M} \right)^{\frac{n+1}{n}}
\eea
The minimum ratio ${B_n}/{F_n}$ is achieved at $\varepsilon=0$, corresponding to the maximization of both the entropy $H$ and the Kolmogorov-Sinai entropy $h_{KS}$. Under these conditions, we obtain:
\bea
\ds \frac{B_1}{F_1}=\frac{1}{4}, \ \ \ \lim_{n \rightarrow \infty}\left( \frac{B_n}{F_n}\right) = \frac{1}{2}.
\eea
Given that ${B_n}/{F_n}$ is a monotonically increasing sequence, it can be inferred that the backward-to-forward scattering ratio is confined within the interval $[0.25, 0.5)$, depending on the specific scale under consideration. While local variations may arise depending on the integration bounds of $\Lambda_L$, the fact that ${B_n}/{F_n}$ remains bounded by non-zero values proves that backscattering is an intrinsic component necessary for a proper description of the energy cascade as a combined "stretch-and-fold" mechanism in the presence of incompressibility. Such confinement suggests the existence of a geometric universality in turbulence that transcends specific wavenumbers.
While the classical Eulerian framework analyzes nonlinear triadic interactions in the Fourier space of the Navier-Stokes equations to study inter-scale kinetic energy transfer, the present Lyapunov-Lagrangian approach -found to be equivalent to the Eulerian description \cite{Truesdell77}- focuses on fluid deformation across scales. Consequently, the nonlinear triadic interactions are here mapped into stretch-and-fold processes characterized by Lagrangian Lyapunov analysis. This offers a significant advantage: the complex triadic summations of the Eulerian approach are effectively distilled into a non-zero mean PDF, $F_\Lambda$, which naturally incorporates both positive and negative values of $\Lambda_L$. If Eulerian triads describe how turbulent energy flows between scales, Lagrangian Lyapunov exponents quantify the extent to which the flow is engaged in the direct cascade (forward) or in the reorganization of coherent structures (backscatter) through the constraint of incompressibility.

While the framework developed in this work is fundamentally theoretical, its physical predictive power is directly verifiable through a comparison with standard numerical experiments. Crucially, the analytical thresholds derived from our Lyapunov–Liouville trajectory analysis—predicting that forward and backward scattering events occur with exact probabilities of \(2/3\) and \(1/3\), respectively—find an immediate and robust quantitative validation in state-of-the-art Direct Numerical Simulations (DNS) of isotropic turbulence. As documented in recent high-resolution DNS studies (see, e.g., \cite{Yao2024} and \cite{Cardesa}), statistical partitioning of the scale-to-scale energy flux in the inertial range yields a conditional probability ratio for signed cascade events that asymptotically converges to \(\approx 2.0\). The fact that an independent, purely numerical DNS dataset recovers the exact fractional values generated by our parameter-free stochastic closure confirms that the Hänggi–Klimontovich post-point representation captures the genuine, endogenous physics of chaotic Lagrangian advection in isotropic flows.

\bigskip

\section{Most Probable ratio between Lagrangian Lyapunov exponents}
\begin{figure}[h!]
	\centering
	\includegraphics[width=110mm,height=90mm]{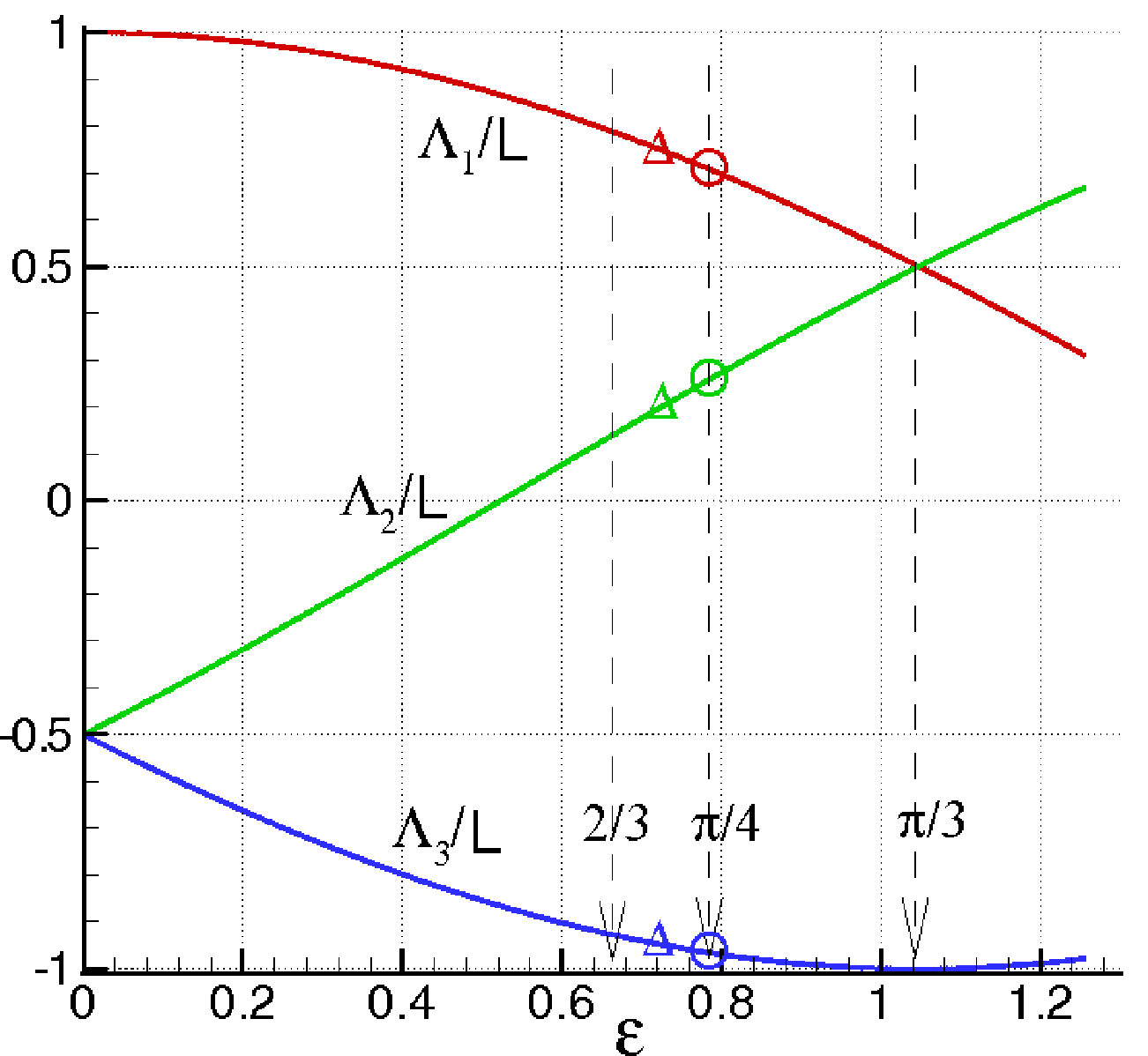}
\caption{Lagrangian Lyapunov exponents in function of the phase $\varepsilon$.}
\label{figura_1}
\end{figure}

To obtain the most probable ratios between the Lagrangian Lyapunov exponents, which are representative of homogeneous isotropic turbulence, consider a generic Lyapunov vector evolving according to the law:
\bea
\ds {\bfxi} = \Sum_{k=1}^3 {\bf e}_k(t) \eta_k \exp(\Lambda_L^{(k)} t)\equiv
\exp(\Lambda_L^{(1)}t) \Sum_{k=1}^3 {\bf e}_k(t) \eta_k \exp((\Lambda_L^{(k)} -\Lambda_L^{(1)})t)
\label{lyap vect}
\eea
where ${\bf e}_k(t)$ are the unit vectors of the Lyapunov basis, which changes orientation according to the fluid motion, and $\eta_k$ are the corresponding components, represented by slowly varying functions of time. Thus, $\Lambda_L^{(1)} \ge \Lambda_L^{(2)} \ge \Lambda_L^{(3)}$, with $\Lambda_L^{(1)} =L, \Lambda_L^{(2)} = \Lambda_L^{(3)}=-L/2$. Equation (\ref{lyap vect}) expresses also the Ott's theorem \cite{Ott2002}, which states that $\bfxi$ tends to align with the direction of maximum growth ${\bf e}_1(t)$. Given the hypothesis of isotropic turbulence, the initial conditions correspond to $\eta_1=\eta_2=\eta_3$. Consequently, the time required for the second and third terms to decay to the value $1/e$ is $2/3/L$. Furthermore, with reference to Eq. (\ref{Lambda}), a phase variation $\Delta \varepsilon = \pi/3$ corresponds to a discontinuous rotation of the Lyapunov basis--i.e., the maximal growth direction undergoes a jump of $\pi/2$.
This specific time scale is representative for evaluating the proportions between the Lyapunov exponents observed in isotropic turbulence. To identify these exponents, a phase $\varepsilon \in (2/3, \pi/3)$ must be selected for use in Eq. (\ref{Lambda}). Although the $\Lambda_L^{(k)}$ fluctuate continuously due to the bifurcations of the velocity gradient, an estimate of these exponents--for instance, for $\varepsilon \in(0.71, \pi/4)$--should provide a valid reference for the mean ratios observed in isotropic turbulence. In particular, the value $\varepsilon \simeq 0.71$ can be reasonably suitable for representing instantaneous or finite-time exponent ratios, while $\varepsilon= \pi/4$ should yield values closer to the asymptotic Lyapunov exponents. According to this criterion, the most probable ratios in homogeneous isotropic turbulence are given by:
\bea
\begin{array}{l@{\hspace{-0.cm}}l}
\ds \Lambda_L^{(1)}: \Lambda_L^{(2)}: \Lambda_L^{(3)} \simeq 4.1: 1.: -5.1 \ 
\mbox{for} \ \varepsilon=0.71, \\\\
\ds \Lambda_L^{(1)}: \Lambda_L^{(2)}: \Lambda_L^{(3)} \simeq 2.75: 1.: -3.75 \ \ \mbox{for} \ \varepsilon=\frac{\pi}{4},
\end{array}
\label{ratios}
\eea 
Figure \ref{figura_1} shows the variations of the three Lyapunov exponents in function of $\varepsilon$, where triangular and circular symbols indicate the ratios given by Eq. (\ref{ratios}).   
These results are in good agreement with numerical simulation data in the literature \cite{Ashurst, nomura1998}, which indicate that the average structures observed in isotropic turbulence correspond to ribbon--like structures. These structures twist around the direction of maximal growth and elongate longitudinally due to the presence of two positive exponents.
 This demonstrates the validity of the results obtained from this analysis, as well as the effectiveness of the  proposed non--diffusive closures.

\bigskip

\section{Eddy Diffusivities and Turbulent Prandtl Number}

This section explores the connection between the proposed closures (\ref{K G}) and (\ref{fluc K G}), derived in previous work \cite{deDivitiis2026a}, and the classical framework of eddy diffusivities. This link demonstrates that eddy diffusivities can be interpreted as fluctuating quantities governed by a specific stochastic process. Their fluctuations follow those of the Lagrangian Lyapunov exponent through stochastic differential equations, while their mean magnitudes are directly coupled to the velocity and temperature correlation functions.

In accordance with the formal structure of the Navier-Stokes and heat equations, we define the fluctuating eddy diffusivities $\tilde{\nu}_T$ and $\tilde{\kappa}_T$ as the quantities satisfying the following relations:
\bea
\begin{array}{l@{\hspace{-0.cm}}l}
\ds \tilde{K} =  \Lambda_L r \ u^2 \frac{\partial f}{\partial r}=2 u^2 
 \frac{1}{r^4} \frac{\partial}{\partial r} \left( r^4 \tilde{\nu}_T \frac{\partial f} {\partial r} \right), \\\\
\ds  \tilde{G} =  \Lambda_L r \ \theta^2 \frac{\partial f_\theta}{\partial r}= 2 \theta^2
 \frac{1}{r^2} \frac{\partial}{\partial r} \left( r^2 \tilde{\kappa}_T \frac{\partial f_\theta}{\partial r} \right)
\end{array}
\label{fluc K G nu k}
\eea
Consequently, $\tilde{\nu}_T$ and $\tilde{\kappa}_T$ (and thus $\tilde{K}$ and $\tilde{G}$) inherit the fluctuations of $\Lambda_L$ described by the PDF in Eq. (\ref{Pl}). These fluctuating diffusivities may therefore assume both positive and negative values depending on the fluctuations of $\Lambda_L$, effectively accounting for backscattering and forward scattering phenomena, respectively. Equations (\ref{fluc K G nu k}) can be rearranged as:
\bea
\begin{array}{l@{\hspace{-0.cm}}l}
\ds \frac{\partial \tilde{\nu}_T}{\partial r}  
   + \tilde{\nu}_T \left(\frac{4}{r} 
   + \frac{\partial \ln f'}{\partial r}\right)  = \frac{\Lambda_L}{2} \ r, \\\\
\ds \frac{\partial \tilde{\kappa}_T}{\partial r}  
   + \tilde{\kappa}_T \left(\frac{2}{r} 
   + \frac{\partial \ln f'_\theta}{\partial r}\right)  = \frac{\Lambda_L}{2} \ r, 
\end{array}
\label{fluc K G nu k bis}
\eea
where the apex denotes the differentiation with respect to $r$. These are differential equations with a stochastic forcing term $\Lambda_L(r)$, highlighting the direct impact of $\Lambda_L$ fluctuations on both diffusivities. This influence is not manifested through algebraic functions of the instantaneous value of $\Lambda_L$, but rather as a dynamical response mediated by linear differential equations, given the deterministic gradients of the autocorrelation functions. Thus, $\tilde{\nu}_T$ and $\tilde{\kappa}_T$ exhibit fluctuations corresponding to the response of Eqs. (\ref{fluc K G nu k bis}) to the stochastic input $\Lambda_L$ governed by the PDF in Eq. (\ref{Pl}), i.e.
\bea
\begin{array}{l@{\hspace{-0.cm}}l}
\ds \tilde{\nu}_T(r)=\frac{1}{2 f'(r) r^4} \int_0^r \Lambda_L(s) s^5 f'(s) \ ds, \\\\
\ds \tilde{\kappa}_T(r)=\frac{1}{2 f'_\theta(r) r^2} \int_0^r \Lambda_L(s) s^3 f'_\theta(s) \ ds.
\end{array}
\eea

The mean values $K$ and $G$, as well as the mean diffusivities $\nu_T = \langle \tilde{\nu}_T\rangle_L$ and $\kappa_T = \langle \tilde{\kappa}_T \rangle_L$, are obtained by averaging Eqs. (\ref{fluc K G nu k}) using the PDF (\ref{Pl}), which provides the statistics of the Lagrangian Lyapunov exponent:
\bea
\begin{array}{l@{\hspace{-0.cm}}l}
\ds K(r) = u^3 \sqrt{\frac{1-f}{2}} \frac{\partial f}{\partial r}=2 u^2 
 \frac{1}{r^4} \frac{\partial}{\partial r} \left( r^4 \nu_T \frac{\partial f} {\partial r} \right), \\\\
\ds  G(r) = u \theta^2 \sqrt{\frac{1-f}{2}} \frac{\partial f_\theta}{\partial r}=2 \theta^2
 \frac{1}{r^2} \frac{\partial}{\partial r} \left( r^2 \kappa_T \frac{\partial f_\theta}{\partial r} \right)
\end{array}
\label{K G n k}
\eea

To establish a pragmatic connection with established literature, we express the velocity and temperature autocorrelations, at least locally, via power laws with scaling exponents $m$ and $n$:
\bea
\begin{array}{l@{\hspace{-0.cm}}l}
\ds f(r) \simeq 1-a r^m, \\\\
\ds f_\theta(r) \simeq 1-b r^n,
\end{array}
\label{auto_corr}
\eea
where $m=n=2/3$ describes the Kolmogorov inertial range. Regarding the fluctuating diffusivities $\tilde{\nu}_T$ and $\tilde{\kappa}_T$, substituting Eqs. (\ref{auto_corr}) into Eqs. (\ref{fluc K G nu k bis}) yields the following stochastic differential equations in terms of $m$ and $n$:
\bea
\begin{array}{l@{\hspace{-0.cm}}l}
\ds \frac{\partial \tilde{\nu}_T}{\partial r}  
   + \left( 3+m \right)  \frac{\tilde{\nu}_T}{r} 
    = \frac{\Lambda_L}{2} \ r, \\\\
\ds \frac{\partial \tilde{\kappa}_T}{\partial r}  
   + \left(1+ n \right) \frac{\tilde{\kappa}_T}{r} 
    = \frac{\Lambda_L}{2} \ r, 
\end{array}
\label{fluc K G nu k ter}
\eea

Concerning the mean eddy diffusivities, $\nu_T$ and $\kappa_T$, the substitution of Eqs. (\ref{auto_corr}) into Eqs. (\ref{K G n k}) results in:
\bea
\begin{array}{l@{\hspace{-0.cm}}l}
\ds \nu_T = u  \sqrt{\frac{a}{2}}  \ \frac{r^{1+m/2}}{8 + 3 m}, \\\\
\ds \kappa_T = u \sqrt{\frac{a}{2}} \ \frac{r^{1+m/2}}{4+  m + 2 n}, 
\end{array}
\eea
These relations can be expressed in terms of the second-order velocity structure function $F_2$, which in isotropic turbulence is given by:
\bea
\ds F_2 = \left\langle \left( {\bf u}({\bf x}+{\bf r}) - {\bf u}({\bf x})\right)  \cdot \left( {\bf u}({\bf x}+{\bf r}) - {\bf u}({\bf x})\right) \right\rangle = 6 u^2 \left( 1-f -\frac{1}{3} \frac{\partial f}{\partial r} r\right) 
\eea
This allows us to explicit $\nu_T$ and $\kappa_T$ in terms of $m$, $n$, and $F_2$:
\bea
\begin{array}{l@{\hspace{-0.cm}}l}
\ds \nu_T = \frac{F_2^{1/2} r}{2 \sqrt{3+ m} \ (8+3m)}, \\\\
\ds \kappa_T =  \frac{F_2^{1/2} r}{2 \sqrt{3+ m} \ (4+m + 2 n)}, 
\end{array}
\label{n k F_2}
\eea
\begin{figure}[h!]
	\centering
	\includegraphics[width=95mm,height=99mm]{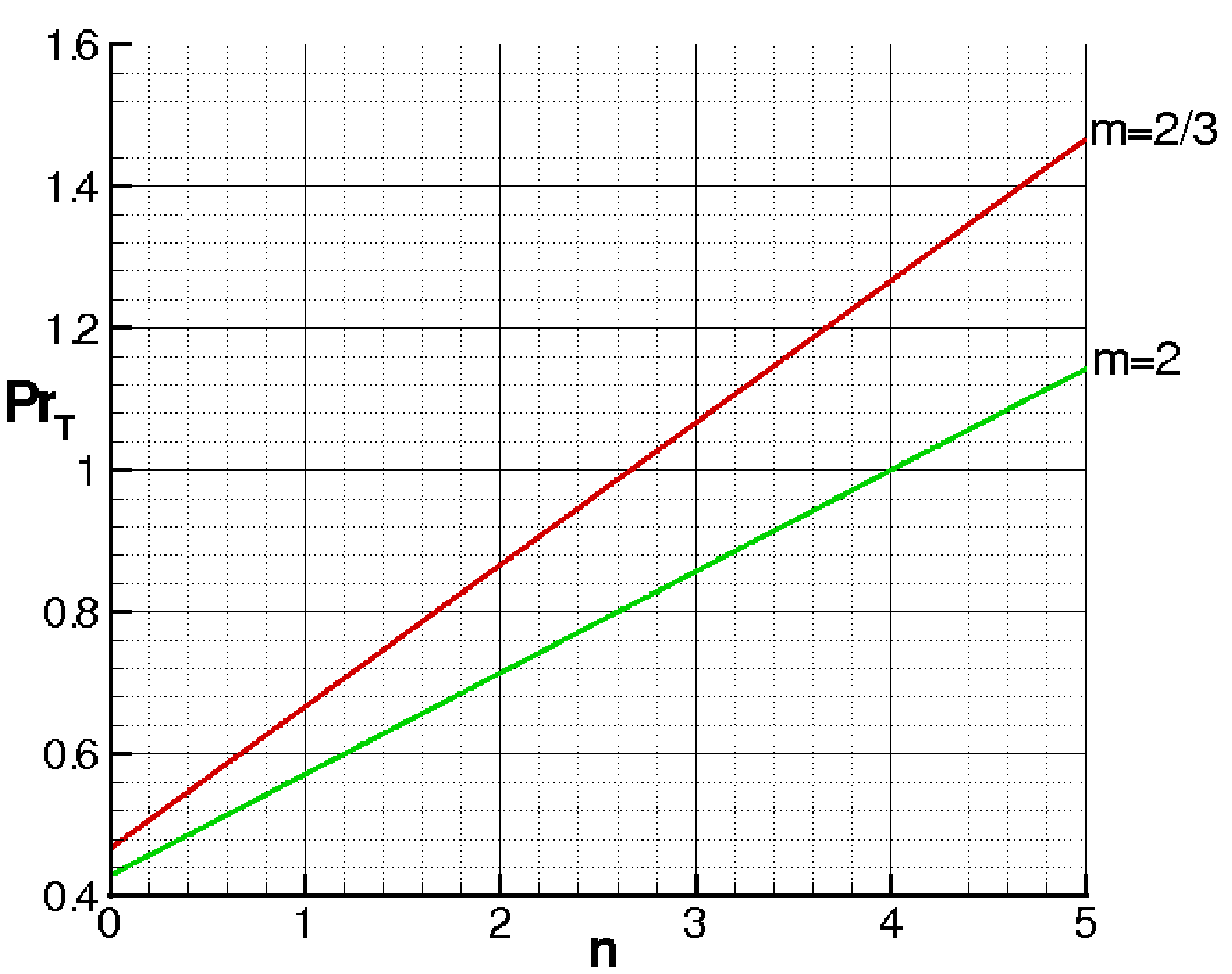}
\caption{Turbulent Prandtl number $Pr_T$ in terms of temperature correlation scaling exponent $n$, for inertial and viscous regimes.}
\label{figura_2}
\end{figure}
It is crucial to highlight that the analytical expression for the eddy viscosity derived in Eq.  (\ref{n k F_2})  identically recovers the mathematical architecture of the classical Structure-Function (SF) subgrid-scale model introduced by M\'etais and Lesieur (1992) \cite{Lesieur1996}. In their pioneering framework, the subgrid-scale eddy viscosity is formulated directly in physical space to account for local turbulence intermittency, scales with the grid mesh \(\Delta x\), and depends on the local second-order velocity structure function \(F_2(\mathbf{x}, \Delta x)\) according to \(\nu_T \propto \Delta x [F_2(\mathbf{x}, \Delta x)]^{1/2}\). However, a major conceptual and predictive distinction lies in how the scaling prefactors are determined. The formulation in Ref. \cite{Lesieur1996} computes the numerical prefactor (analytically evaluated as \(0.105 \, C_K^{-3/2}\)) by invoking Batchelor’s relation, which requires prescribing a  Kolmogorov spectrum up to the cutoff wavenumber \(k_c = \pi/\Delta x\). Consequently, its magnitude remains strictly bound to the empirical choice of the Kolmogorov constant \(C_{K}\). Conversely, the scaling coefficients in our Eq. (\ref{n k F_2}) are derived from a closed, first-principles approach. Rather than relying on empirical constants, the parameters emerge self-consistently from the Lagrangian trajectory stability analysis, dictated exclusively by the volume-preserving constraints of the Lyapunov–Liouville theorem. Equation  (\ref{n k F_2}) therefore provides a purely dynamic, parameter-free theoretical foundation that justifies the empirical success of the structure-function model directly from the chaotic internal structure of the fluid.

This confirms the validity of the Liouville--Lyapunov analysis regarding the proposed closures. In particular, in the Kolmogorov regime (\(m=2/3\)), the constant for \(\nu _{T}\) is remarkably close to the value reported in \cite{Lesieur1996}.  Specifically, adopting the Kolmogorov coefficient obtained in \cite{deDivitiis2026a} through the present non-diffusive closures, the prefactor from Lesieur \cite{Lesieur1996} is 0.02625, whereas the present theory yields 0.02611.

  We can then determine the turbulent Prandtl number as $Pr_T=\nu_T/\kappa_T$:
\bea
\ds Pr_T= \frac{\nu_T}{\kappa_T} = \frac{4+m + 2 n}{8+3m}
\eea
According to the present analysis, $Pr_T$ depends on $m$ and $n$ alone, in the sense that in intervals of $r$ where the scaling exponents are given, the turbulent Prandtl number does not change.  Evaluating $Pr_T$ in the Kolmogorov ($m=2/3$) and viscous ($m=2$) regimes of the energy spectrum yields the two linear characteristics shown in Fig. \ref{figura_2}. While $Pr_T$ increases with $n$, it remains confined between 0.4 and 1.5. Specifically, $n$ depends on the molecular Prandtl number $Pr = \nu/\kappa$ and the relevant scaling region of the temperature spectrum. For $Pr \approx 1$, the temperature spectrum mirrors the kinetic energy spectrum, and in the inertial regime ($n=m=2/3$), we obtain $Pr_T=0.6$.

Conversely, values of $n$ approaching zero correspond to cases where $Pr \gg 1$ (e.g., oils), where the temperature correlation is characterized by a logarithmic scaling law (viscous-convective range $\Theta(k) \approx k^{-1}$) at scales smaller than the Kolmogorov scale \cite{Batchelor_2}; this implies $Pr_T \simeq 0.47$. On the other hand, relatively high values of $n$ correspond to $Pr \ll 1$, where the high thermal conductivity relative to kinematic viscosity produces significantly different conditions \cite{Batchelor_3}. In the inertial regime, scaling laws such as $\Theta(k) \approx k^{-11/3}$ or $\Theta(k) \approx k^{-17/3}$ \cite{Rogallo, deDivitiis_4} yield values of $n$ slightly less than 3 and 5, respectively, leading to $Pr_T$ values considerably higher than unity, as illustrated in Fig. \ref{figura_2}.

\bigskip

\section{Conclusion \label{Conclusion}}

The primary contribution of this study lies in the quantitative evaluation of forward and backward scattering from a Lagrangian perspective. This is achieved through the formal derivation of the Lagrangian Lyapunov exponent distribution by modeling the "stretch and fold" mechanism as a \textbf{H\"anggi--Klimontovich stochastic process}. Its transformation into an equivalent \textbf{It\^o SDE} reveals how the \textbf{spurious drift}, here interpreted as the alignment dynamics of Lyapunov vectors, balances stochastic dispersion to determine the statistical evolution of the system.  The resulting \textbf{Fokker--Planck formulation} provides a rigorous dynamical foundation for the uniform quasi-PDF of $\Lambda_L$, where the continuous nature of this distribution is fundamentally driven by a Lagrangian bifurcation rate significantly higher than the Lyapunov exponents themselves ($\sigma \ggg \Lambda_L$). This confirms that the turbulent state emerges from a precise balance between deterministic stretching and high-frequency stochastic redistribution.

This dynamical description is shown to be fundamentally consistent with a principle of \textbf{maximum entropy}. Specifically, it has been demonstrated that the analytical closures for the von K\'arm\'an--Howarth and Corrsin equations correspond to the simultaneous maximization of the \textbf{information entropy} associated with the Lyapunov distribution and the \textbf{Kolmogorov--Sinai entropy}. This statistical-mechanical consistency suggests that the evolution of turbulence is governed by the optimal dispersion of Lagrangian trajectories in phase space, providing a robust physical justification for the emergence of backscattering effects.

Furthermore, the proposed framework enables a precise quantitative assessment of forward and backward scattering. Within this formulation, forward scattering is directly linked to the stretching of Lyapunov vectors and trajectory instabilities, while backscattering is shown to emerge from the constraints of fluid incompressibility. The theoretical ratio between backscattering and forward scattering derived from the uniform PDF of $\Lambda_L$ is found to be in close agreement with numerical and experimental benchmarks available in the literature. By leveraging the alignment properties of Lyapunov vectors, the theory successfully predicts the observed canonical Lyapunov exponents and fundamental transport properties, such as eddy viscosity and the turbulent Prandtl number. These parameters, while traditionally associated with diffusive models, are here shown to emerge naturally from the underlying \textbf{non-diffusive Lagrangian dynamics}. In conclusion, the integration of H\"anggi--Klimontovich stochastics with maximum entropy principles establishes a comprehensive bridge between microscopic trajectory instabilities and the macroscopic statistical properties of isotropic turbulence, providing a full quantitative characterization of both forward and backward scattering, as well as the related eddy diffusivities.

\bigskip 

\section{Acknowledgments}

This work was partially supported by the Italian Ministry for the Universities 
and Scientific and Technological Research (MIUR). 

\bigskip

\end{document}